\documentclass[preprint,amsmath,aps,amssymb,nofootinbib,superscriptaddress,floatfix]{revtex4-2}
\usepackage{bm,amssymb,amsmath,siunitx,color,booktabs}
\usepackage[justification=raggedright,singlelinecheck=true]{caption}
\usepackage[mathscr]{euscript} 
\usepackage{graphicx}
\usepackage{upgreek}
\usepackage{dcolumn}
\usepackage{multirow}
\usepackage{bm}
\usepackage{url}
\usepackage{float}
\usepackage{caption}
\usepackage{subcaption}
\usepackage{makecell}
\usepackage{amsmath}
\usepackage{booktabs}
\usepackage{listings}
\usepackage{threeparttable}

\begin{document}

\title{A Dynamical Equilibrium Linking Nanohertz Stochastic Gravitational Wave Background to Cosmic Structure Formation}

\author{Manjia Liang}
\email{liangmanjia21@mails.ucas.ac.cn}
\affiliation{Center for Gravitational Wave Experiment, National Microgravity Laboratory, Institute of Mechanics, Chinese Academy of Sciences, Beijing 100190, China}

\author{Peng Xu}
\email{xupeng@imech.ac.cn}
\affiliation{Center for Gravitational Wave Experiment, National Microgravity Laboratory, Institute of Mechanics, Chinese Academy of Sciences, Beijing 100190, China}
\affiliation{Taiji Laboratory for Gravitational Wave Universe (Beijing/Hangzhou), UCAS, Beijing 100049, China}
\affiliation{School of Fundamental Physics and Mathematical Sciences, Hangzhou Institute for Advanced Study, UCAS, Hangzhou 310024, China}
\affiliation{Lanzhou Center of Theoretical Physics, Lanzhou University, Lanzhou 730000, China}

\author{Ruijun Shi}
\affiliation{School of Physics and Astronomy, Beijing Normal University, Beijing, 100875, China}

\author{Zhoujian Cao}
\affiliation{School of Physics and Astronomy, Beijing Normal University, Beijing, 100875, China}
\affiliation{School of Fundamental Physics and Mathematical Sciences, Hangzhou Institute for Advanced Study, UCAS, Hangzhou 310024, China}

\author{Ziren Luo}
\affiliation{Center for Gravitational Wave Experiment, National Microgravity Laboratory, Institute of Mechanics, Chinese Academy of Sciences, Beijing 100190, China}
\affiliation{Taiji Laboratory for Gravitational Wave Universe (Beijing/Hangzhou), UCAS, Beijing 100049, China}
\affiliation{School of Fundamental Physics and Mathematical Sciences, Hangzhou Institute for Advanced Study, UCAS, Hangzhou 310024, China}

\author{Minghui Du}
\affiliation{Center for Gravitational Wave Experiment, National Microgravity Laboratory, Institute of Mechanics, Chinese Academy of Sciences, Beijing 100190, China}

\author{Qiong Deng}
\affiliation{Center for Gravitational Wave Experiment, National Microgravity Laboratory, Institute of Mechanics, Chinese Academy of Sciences, Beijing 100190, China}
\affiliation{Lanzhou Center of Theoretical Physics, Lanzhou University, Lanzhou 730000, China}

\author{Bo Liang}
\affiliation{Center for Gravitational Wave Experiment, National Microgravity Laboratory, Institute of Mechanics, Chinese Academy of Sciences, Beijing 100190, China}

\author{Jiaxiang Liang}
\affiliation{Center for Gravitational Wave Experiment, National Microgravity Laboratory, Institute of Mechanics, Chinese Academy of Sciences, Beijing 100190, China}
\affiliation{Taiji Laboratory for Gravitational Wave Universe (Beijing/Hangzhou), UCAS, Beijing 100049, China}

\begin{abstract}
The stochastic gravitational wave background (SGWB) is conventionally treated as a passive relic of its astrophysical and cosmological sources, with negligible back-reaction on the matter content of the Universe. Here we show that this assumption needs to be modified once the SGWB and matter are treated as a dynamically coupled non-equilibrium system. Combining linearized general relativity with the fluctuation-dissipation theorem, we derive a generalized Langevin framework that drives the coupled system toward a dynamical equilibrium, which is characterized by a distinctive strain spectrum with a high-frequency cutoff $\mathcal{W}$, and a scale-dependent coupling parameter that screens gravity progressively for the most massive structures. Three findings support this framework. Fitting the equilibrium spectrum to the NANOGrav 15-year dataset yields a Bayes factor of $48\pm 3.8$ over the supermassive black hole binary baseline, achieved entirely within general relativity and the Standard Model. The PTA-calibrated screening mass scale $m_{c}\sim 10^{12}\text{--}10^{14}\,M_{\odot}$ overlaps, with no free cosmological parameter, the $\Lambda$CDM-derived linear-to-nonlinear transition mass $M_{\rm NL}$ of cosmic structure at $\sim 8\,h^{-1}\,\mathrm{Mpc}$. Most strikingly, promoting this concordance to a structural identification expresses $\mathcal{W}$ entirely in terms of $M_{\rm NL}$, and its inverse acquires a transparent physical reading as a coherence threshold for SGWB-matter coupling. $\mathcal{W}$ is thereby a derived quantity linking nanohertz gravitational-wave observables to the late-time cosmological sector. The framework makes distinctive scale-dependent predictions testable by forthcoming large-scale structure surveys and space-borne gravitational-wave observatories.
\end{abstract}

\maketitle

\section{Introduction}

Modern precision cosmology stands as one of the great intellectual achievements of contemporary physical science. 
The standard cosmological model, $\Lambda$CDM, 
has been remarkably successful in accounting for the cosmic microwave background anisotropies, the accelerating expansion of the Universe, the primordial abundances of light elements, and the emergence of large-scale structure from primordial perturbations~\cite{Baumann2022,Planck2018,Peebles1980,cyburt2016big,peebles2003cosmological}. 
However, despite these triumphs, $\Lambda$CDM confronts a growing list of theoretical challenges and observational tensions. The physical origin of dark energy and dark matter remains unknown; the mechanism driving cosmic inflation lacks direct empirical confirmation; and persistent discrepancies have emerged between independent measurements of key cosmological parameters, most notably the Hubble tension ~\cite{Riess2022,Freedman2021} and the $\sigma_8$ tension between the ones inferred from the cosmic microwave background (CMB) ~\cite{Planck2018} and those determined by low-redshift probes ~\cite{KiDS1000,DESY3}. 
These tensions may point to new physics beyond $\Lambda$CDM, or to systematic effects not yet fully understood. 
Amid this, one piece of such grand landscape has been conspicuously overlooked in the standard structure-growth framework, that the stochastic gravitational wave background (SGWB). Gravitational waves (GW) are regarded as coupling so weakly to matter that they propagate through cosmic structures transparently, exerting no appreciable back-reaction on the matter distribution~\cite{Christensen2018,Kakkat2024}. 
In this work, we re-examine this assumption in the context of cosmic evolution from the perspective of non-equilibrium statistical mechanics and present evidence that the SGWB may play a more active role than has been recognized.

The existence of a low-frequency SGWB has now been firmly established by observations. In 2023, different pulsar timing array (PTA) collaborations, including NANOGrav~\cite{NG15evidence}, the European PTA~\cite{EPTA2023}, the Parkes PTA~\cite{PPTA2023}, and the Chinese PTA~\cite{CPTA2023}, independently reported strong evidence for a stochastic GW signal in the nanohertz band, exhibiting the Hellings--Downs (HD) angular correlations \cite{agazie2023nanograv} characteristic of a GW origin. The physical source of this background remains an open question. The leading astrophysical interpretation attributes it to the incoherent superposition of GWs emitted by a cosmic population of supermassive black hole binaries (SMBHBs)~\cite{Abbott2016,Allen1988}. Alternatively, a variety of cosmological mechanisms have been proposed, including scalar-induced gravitational waves (SIGW) generated during the radiation era, first-order phase transitions (FOPT) in the early Universe, cosmic string networks, inflationary gravitational waves (IGW), and domain wall (DW) decay~\cite{agazie2023nanograv,Afzal2023,Xue2021,Figueroa2024}. Among these, SIGW models currently receive the strongest empirical support from PTA data analyses~\cite{agazie2023nanograv,Afzal2023,Cai2023}. However, none of these models addresses the dynamical interplay between the SGWB and the matter content of the Universe, and they treat the SGWB as a fossil record of its sources rather than as an active participant in cosmic evolution.

Here, we introduce a fundamentally different perspective. 
By combining non-equilibrium statistical mechanics with linearized tensor perturbations in general relativity, we construct a generalized Langevin framework in which the SGWB acts as a stochastic driver of celestial-system motion, while the gravitational radiation emitted by the resulting accelerated motion constitutes a back-reaction that transfers energy from matter distribution back into the GW background. Although this back-reaction is imperceptibly weak on dynamical timescales, its cumulative effect over the evolution of the Universe is substantial and must not be overlooked. 
Since fluctuation and dissipation are not independent, the coupled system of matter and SGWB naturally evolves toward a state which satisfies a fluctuation-dissipation relation~\cite{Kubo1966} and imposes constraints on the equilibrium power spectrum of the SGWB. 

The central result is that this coupling is not uniformly weak, rather, it is strongly scale-dependent. At low frequencies, equivalently on large spatial scales, massive structures couple efficiently to the SGWB and the system evolves toward a dynamical equilibrium in which the energy injected into matter from the GW background is balanced by the energy dissipated back through gravitational radiation. 
At high frequencies, or equivalently for small-scale structures of modest mass, the coupling remains negligibly weak, recovering the conventional picture in which matter distribution is effectively transparent to GWs. This scale-dependent transition from equilibrium to weak coupling arises naturally from the non-Markovian character of the dissipation kernel. It represents a fundamental correction to the longstanding view that GWs propagate through matter fields only transparently, and opens an entirely new picture in which the SGWB actively participates in the evolution of cosmic structure.

We demonstrate that this scheme is strongly supported by current observational data on three independent fronts.
First, fitting the equilibrium strain power spectrum predicted by our model to the NANOGrav 15-year dataset yields a Bayes factor of $48 \pm 3.8$ relative to the standard SMBHB model, comparable to the leading SIGW scenario but achieved without invoking any new physics beyond general relativity and the Standard Model of particle physics.
Second, the PTA data simultaneously fix a characteristic mass scale over which the scale-dependent screening becomes effective, $m_{c}\sim 10^{12}\text{--}10^{14}\,M_{\odot}$, which overlaps with the mass at which the observed Universe transitions from linear to nonlinear gravitational clustering at $\sim 8\,h^{-1}\,\mathrm{Mpc}$~\cite{Planck2018,Tully2008,Scrimgeour2016}, a concordance arising with no free cosmological parameter.
Third, and most strikingly, promoting this concordance to a structural identification with the $\Lambda$CDM-derived nonlinear mass scale $M_{\rm NL}$ yields a direct relation between the DEGWB cutoff and the cosmological sector,
\begin{equation}\label{W_cosmo_intro}
\mathcal{W} \;=\; \frac{c^{3}}{4G\,M_{\rm NL}},
\end{equation}
in which $\mathcal{W}$ is expressed entirely in terms of late-time cosmological parameters. The associated timescale $1/\mathcal{W}$ admits a transparent physical reading as a phase-coherence threshold for SGWB-matter coupling, set by the Schwarzschild light-crossing time of the critical structure scale. $\mathcal{W}$ is thereby not a fundamental or Planckian scale, nor a regularization imposed by hand, but a derived quantity linking nanohertz gravitational-wave observables to the primordial power spectrum, matter content, and expansion rate of the Universe. The $\Lambda$CDM and DEGWB descriptions of this boundary are complementary rather than mutually exclusive, and their structural agreement constitutes a cross-sector consistency test of the framework. Because $\mathcal{W}$ is fixed entirely by the PTA data with no additional cosmological parameter, the model makes a distinctive prediction that structure growth is suppressed only for the most massive aggregates, with standard behavior recovered on smaller scales, a pattern that can be tested by forthcoming large-scale structure surveys.

\section{Theory}

The physical picture underlying our model rests on a simple observation: any celestial system immersed in a stochastic gravitational wave background is subject to two competing influences. The SGWB exerts a fluctuating tidal force that drives random motion, while the accelerated motion radiates gravitational energy back into the background, producing dissipation. These two processes, fluctuation and dissipation, are not independent. Over cosmological timescales, they drive the coupled system of matter and SGWB toward a state of dynamical equilibrium in which the energy injected into matter by the background is, on average, balanced by the energy radiated back. This balance is the gravitational analogue of the equilibrium reached by a charged particle in a blackbody radiation cavity, where the stochastic Lorentz force is balanced by Larmor radiation reaction~\cite{Ford1985,Ford1988}. The key difference is that in gravity, mass plays the dual role of charge and inertia, making the coupling strength inherently mass-dependent, a feature with far-reaching cosmological consequences.

Statistical analyses in curved spacetime are generally subtle~\cite{Earman1978,Hakim1967,Dunkel2009}. The SGWB energy density is small enough that it can be treated as a linear perturbation $h_{\mu\nu}$ over an expanding background, and the stochastic motions it induces are non-relativistic. We therefore adopt the weak-field and slow-motion approximations throughout, and work in the global rest frame with respect to the SGWB (analogous to the CMB rest frame) in which the background is statistically homogeneous and isotropic.

Under these approximations, the fluctuating force on a test mass $m$ is obtained directly from the geodesic equation,
\begin{equation}
\frac{{d}^{2}{x}^{\mu }}{d{s }^{2}}+ {\varGamma }_{\rho \sigma  }^{\mu }\frac{d{x}^{\rho }}{{d s }}\frac{d{x}^{\sigma  } }{{d s }}=0,
\end{equation}
where ${x}^{\mu }=\{t,x^i\},\ (i=1,2,3)$ are the coordinates of the test mass in the cosmological rest frame of the SGWB, ${\varGamma }_{\rho \sigma  }^{\mu }$ represent the connection, and $s$ is the proper time of the test mass. The fluctuating force from tensor perturbations $h_{\mu\nu}$ can be expressed as,
\begin{equation}
F^{i}(t)= \frac{-mc^2 }{2}\left(\eta^{\lambda i} - h^{\lambda i}\right)\left(2h_{\lambda 0,0} - h_{00,\lambda}\right).
\end{equation}
For a statistically homogeneous and isotropic SGWB, this force has zero mean and a power spectral density proportional to $m^{2}\omega^{2}c^2S(\omega)$, where $S(\omega)$ is the strain power spectral density of the SGWB (see the Methods section for detailed definitions and derivations). The corresponding dissipation arises from gravitational radiation back-reaction, that a mass $m$ following a the stochastic zigzag motion 
in the SGWB will interchange energy with the wave field background.
This can be derived in terms of the linearized Einstein equation, please see the Methods section for details.

To describe the interplay between fluctuation and dissipation we adopt the generalized Langevin equation~\cite{Ford1988,Bao2005,Kubo1966,Kubo1991}
\begin{equation}\label{gl}
m\frac{dv^{i}(t)}{dt} = -m\int_{-\infty}^{t}\gamma(t-t')\,v^{i}(t')\,dt' + F^{i}(t),
\end{equation}
where $\gamma(t)$ is the dissipation kernel encoding memory of past interactions and $v^i$ the velocity of the mass. The non-Markovian structure is essential: gravity is a long-range interaction, and the back-reaction at any instant depends on the entire history of the system's motion. 

Given the natural and concise assumptions above, this model will lead to the following important predictions and conclusions (see Methods for details).  

\textit{Why the system approaches dynamical equilibrium.}---For any finite coupling, the fluctuating force continuously injects kinetic energy into the test mass while radiation reaction continuously extracts it. If injection exceeds extraction, increasing $|\vec{a}|^{2}$ and hence the dissipated power, until the two rates balance; conversely, if extraction exceeds injection, the kinetic energy decreases until equilibrium is restored. This negative feedback guarantees a stable fixed point, and the system approaches it whenever the coupling is non-vanishing. 
The approach to equilibrium is also expected to be rapid on cosmological timescales. As shown in the Observational Evidence section, the characteristic relaxation time defined below Eq.~(\ref{dandao}) is calibrated by PTA observations to satisfy $\tau \ll 1/H_{0}$. The coupled SGWB and matter system has therefore had ample time to settle into the dynamical equilibrium by the late time Universe.

\textit{The equilibrium spectrum.}---Imposing the equilibrium condition, energy conservation together with the gravitational radiation-reaction formula (see Methods) constrains the real part of the Fourier transform of the dissipation kernel,
\begin{equation}\label{haosan}
\mathrm{Re}\,\gamma(\omega) = \frac{4Gm\omega^2}{c^3} f_c(\omega),
\end{equation}
where $f_c(\omega) = \mathcal{W}^{2}/(\omega^{2}+\mathcal{W}^{2})$ is a Lorentzian cutoff factor with cutoff frequency $\mathcal{W}$. Combined with the fluctuation-dissipation relation $\mathrm{Re}\,\gamma(\omega) = S_{|\vec{F}|}(\omega)/(6mk_{B}T)$, this determines the equilibrium strain power spectral density of the SGWB,
\begin{equation}\label{yingbian}
S_{h}(\omega) = A^{2}f_c(\omega),
\end{equation}
where $A$ is a constant representing the overall amplitude. The corresponding energy density spectrum ${d{\rho }_{GW} (\omega)}/{d\omega }\propto \omega^{2}S_{h}(\omega)$ takes the form of the Rayleigh-Jeans law, a natural consequence of the classical equipartition theorem applied to a stochastic wave field.

\textit{Why a cutoff is necessary.}---A high-frequency cutoff $\mathcal{W}$ in the equilibrium spectrum is physically required, with both a formal and a physical justification. Formally, without such a cutoff ($\mathcal{W}\to\infty$) the total dissipated power diverges, since every mode of the SGWB contributes to the radiation reaction and the integral over all frequencies is unbounded. Physically, the cutoff has a transparent origin as a coherence threshold for SGWB-matter coupling, set by the Schwarzschild light-crossing time of the critical structure scale, as derived in the Observational Evidence section. The precise functional form of the cutoff---Lorentzian, smooth, or exponential---does not affect the long-time dynamics; only the cutoff frequency $\mathcal{W}$ itself enters the physical predictions (see Table~\ref{tab:2} for observational confirmation of this robustness).


\textit{Scale-dependent coupling and gravitational screening.}---From Eq.~(\ref{haosan}), the dissipation kernel in the time domain is $\gamma(t) = 4Gm\mathcal{W}^{2}[\delta(t) - \mathcal{W}e^{-\mathcal{W}|t|}]/c^3$. The resulting diffusion is ballistic~\cite{Bouchaud1990,Hughes1995}, a direct consequence of the non-Markovian memory. Introducing the non-Markovian factor $b = 1/(1+4Gm\mathcal{W}/c^3)$, the long-time mean velocity under the combined action of the SGWB and a constant external force $\vec{F}_{\mathrm{ext}}$ (e.g.\ from a distant attractor) takes the form
\begin{equation}\label{dandao}
\langle \vec{v}(t\gg\tau)\rangle = b\,\vec{v}(0) + \frac{\vec{F}_{\mathrm{ext}}\,b\,t}{m},
\end{equation}
with relaxation time $\tau = b/\mathcal{W}$. The factor $b$ multiplying $\vec{F}_{\mathrm{ext}}$ reveals the physical mechanism of gravitational screening, that an external gravitational traction is effectively reduced by the factor $b$, or equivalently, gravity acts with an effective coupling
\begin{equation}
G_{\mathrm{eff}}(m) = G\,b = \frac{G}{1+4Gm\mathcal{W}/c^3}.
\end{equation}
The consequences are scale-dependent, that for systems with $4Gm\mathcal{W}/c^3\ll 1$, $b\approx 1$ and $G_{\mathrm{eff}}\approx G$, recovering standard gravitational dynamics; for systems with $4Gm\mathcal{W}/c^3\gtrsim 1$, $b$ decreases, gravitational attraction is screened, and the growth of density perturbations is suppressed. The characteristic mass scale at which this transition sets in is $m_{c}\sim c^3/(4G\mathcal{W})$.
We also stress that this scale-dependent picture is meaningful only within the small-perturbation regime in which our generalized Langevin framework is derived. Large departures of $G_{\mathrm{eff}}$ from $G$ would signal a breakdown of the linearized treatment itself, and accordingly, the model should be understood as predicting a mild, progressive screening near and moderately above $m_{c}$ rather than an arbitrarily strong suppression.

\begin{table}[!ht]
\caption{Bayes factors and $\Delta\text{AIC}$
 values for the DEGWB model and competing scenarios, computed relative to the SMBHB baseline.}
    \centering
    \begin{tabular}{lll}
    \toprule
        model & Bayes factors & $\Delta \text{AIC}$   \\
    \midrule
        DEGWB       & 48.30(3.81) & -8.50 \\ 
        SIGW        & 49.52(3.49) & -6.93 \\
        IGW         & 8.76(1.91)  & -6.89 \\
        DW-sm          & 17.24(1.95) & -4.99 \\
        FOPT        & 5.76(1.13)  & -6.97 \\
        \bottomrule
    \end{tabular}
\label{tab:1}
\end{table}

\begin{table}[!ht]
\caption{Bayes factors and $\Delta \text{AIC}$ value for different DEGWB models}
    \centering
    \begin{tabular}{llll}
    \toprule
        model & Bayes factors & $\Delta AIC$ & $\text{log}_{10}\mathcal{W}$  \\
    \midrule
        DEGWB       & 48.30(3.81) & -8.50 & $-8.24^{+0.14}_{-0.35}$ \\ 
        DEGWB-n     & 34.00(2.94) & -6.69 & $-8.37^{+0.48}_{-1.02}$\\
        DEGWB-exp   & 30.76(2.04) & -9.15 & $-8.24^{+0.20}_{-0.13}$\\
        DEGWB-nocut & 1.39(0.13)  & 0.47  & $\quad --$ \\
        DEGWB-hard cutoff & 8.28(1.58)  & -8.76 & $\quad --$ \\
        \bottomrule
    \end{tabular}
\label{tab:2}
\end{table}

\section{Observational Evidence}


The NG15 data release provides strong Bayesian evidence for the existence of an SGWB in the nanohertz frequency band~\cite{agazie2023nanograv}. 
To assess whether our model can account for this observed background, we compute the Bayes factors and Akaike Information Criteria (AIC) of our DEGWB (Dynamical Equilibrium Gravitational Wave Background) model and other competing scenarios relative to the SMBHB model ~\cite{Figueroa2024}.

\begin{figure}
    \centering
    \includegraphics[width=1\linewidth]{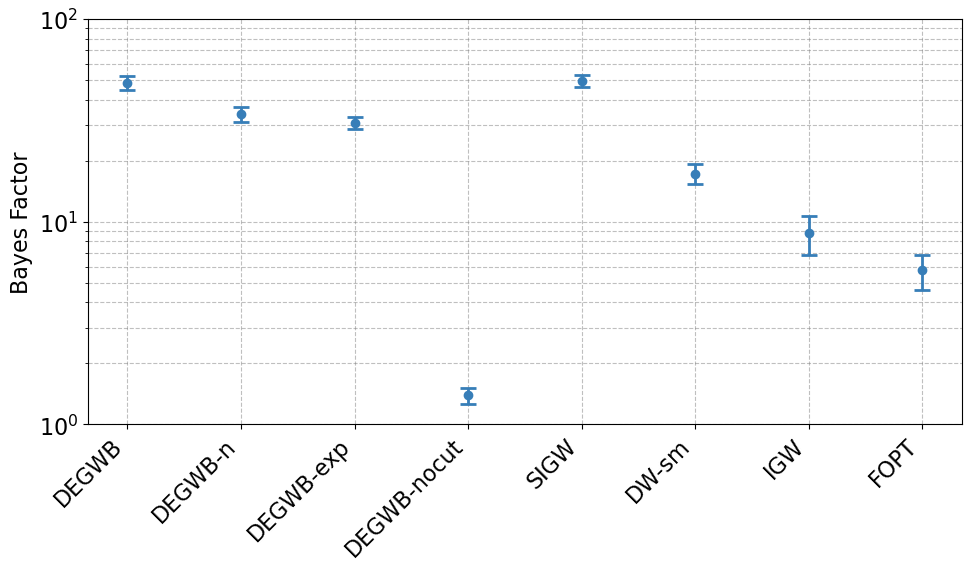}
    \caption{Bayes factors for the model comparisons between the DEGWB, DEGWB with varying cut-off factors, and other physical models. The models' prior distributions are shown in Table \ref{tab:1}.
}
    \label{fig:bf}
\end{figure}

\begin{figure}
    \centering
    \includegraphics[width=0.6\linewidth]{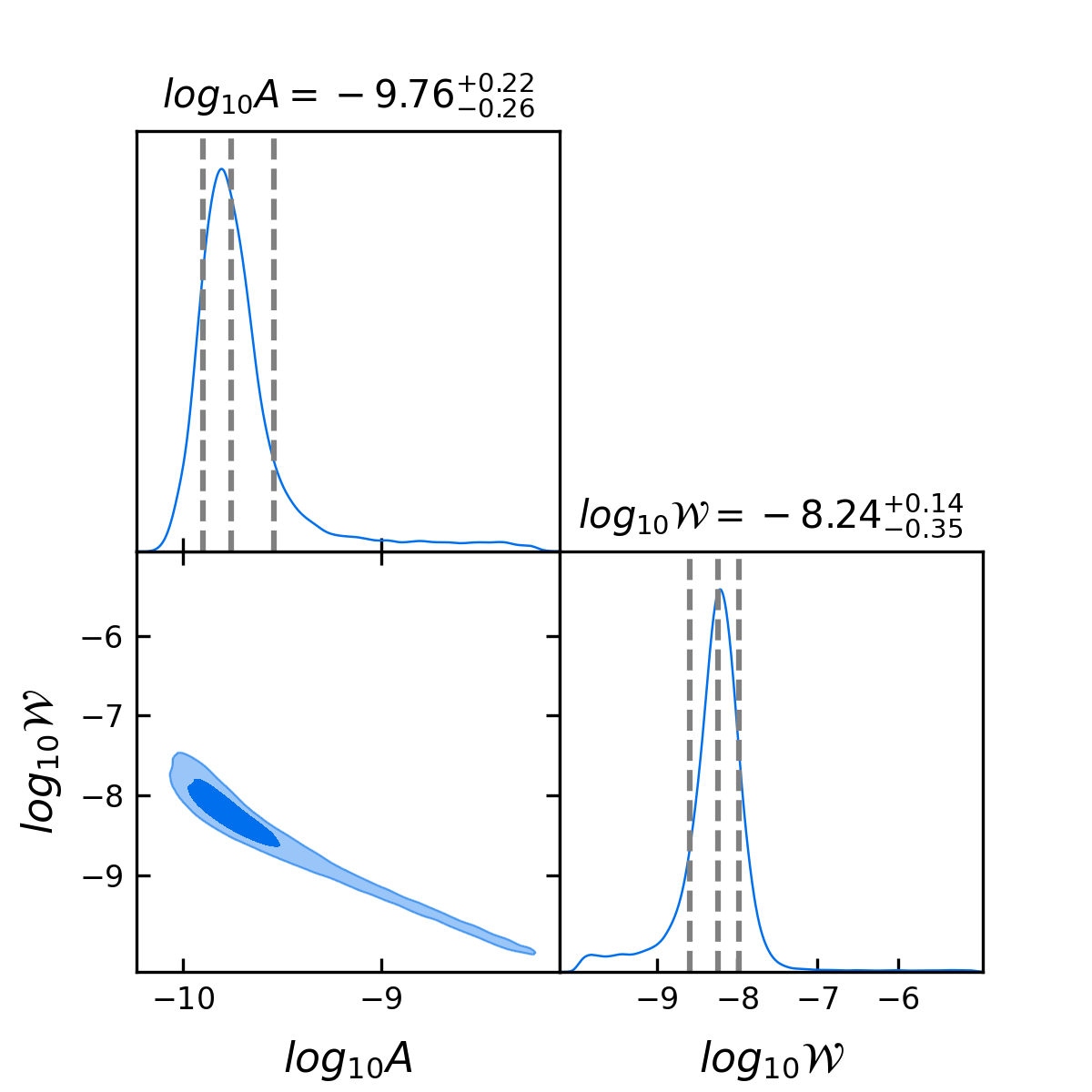}
    \caption{The two-dimensional marginalized posterior distribution of $\log_{10} A$ and $\log_{10}\mathcal{W}$ (for Lorentz cutoff function) with the NANOGrav 15-year dataset.}
    \label{fig:post}
\end{figure}

The resulting model comparisons are summarized in Table~\ref{tab:1}. 
In our analysis, we employ nested sampling to compute the Bayesian evidence for each model, using 14 frequency bins for the gravitational wave spectrum ~\cite{Ashton2022}. The likelihood function is constructed using the free spectrum data under HD correlations from the NG15 dataset, in combination with the \texttt{ceffyl} and \texttt{PTArcade}~\cite{Mitridate2023, NANOGrav_ceffyl} frameworks.
The Bayes factor comparing the DEGWB model to the SMBHB, $\mathcal{B} = 48.30\pm3.81$, indicates strong support based on the NG15 data.  
Among all models tested, DEGWB yields a significantly higher Bayes factor than IGW, FOPT, and DW models, and performs slightly below the SIGW scenario. 
The parameters to be estimated in each model and their prior ranges are shown in Table~\ref{tab:3}.
From the AIC results, DEGWB is clearly preferred over all other individual models, reflecting its superior balance between explanatory power and model simplicity. Taken together, these findings suggest that DEGWB has a high probability of dominating the stochastic signal observed in the NG15 dataset. We organize the Bayesian factors of these models in Figure~\ref{fig:bf} to make the comparison more visual. 
The posterior distributions of the two DEGWB parameters constructed via the \texttt{enterprise}~\cite{Ellis2020enterprise, Taylor2021} framework are shown in Figure ~\ref{fig:post}.

\begin{figure*}
    \centering
    \includegraphics[width=1\linewidth]{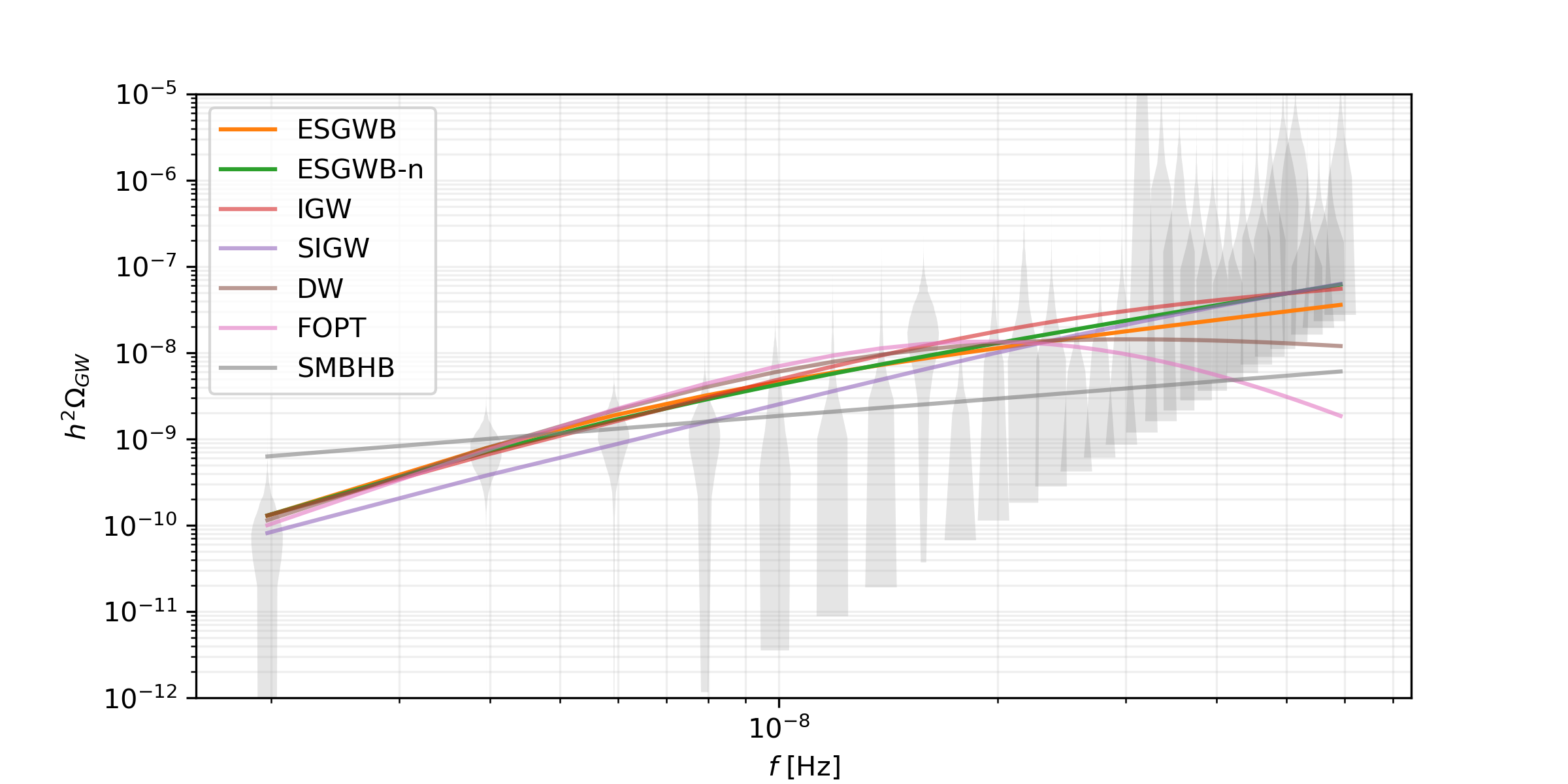}
    \caption{The grey violin plot is the Bayesian free spectrum under Hellings-Downs correlations at frequencies $i/T_{span}$. The violin plot illustrates the posterior distributions of the SGWB energy density across different frequency bins in the NG15 dataset. In the calculation, we use the first 14 bins, $(i_{max}=14)$. The solid lines represent the results of each model under its mean posterior value for the parameters.}
    \label{fig:spet}
\end{figure*}

As emphasized above, our analysis is confined to the classical regime, and we do not attempt to prescribe the specific functional form of the spectral cutoff from physics beyond the Standard Model. The DEGWB should therefore be regarded as a family of models sharing a common, theoretically robust core, that the power-law spectral shape below the cutoff and the physical necessity of a cutoff itself, while differing in the precise functional form through which the cutoff is realized.
To test the robustness of this interpretation, we compare Bayes factors for the DEGWB model under three representative cutoff functions, including the Lorentzian cutoff used above, a smooth cutoff of the form $\mathcal{W}^{n}/(\omega^{n}+\mathcal{W}^{n})$, and an exponential cutoff of the form $(\omega/\mathcal{W})/(e^{\omega/\mathcal{W}}-1)$. The results are summarized in Table~\ref{tab:2}. All three implementations yield large Bayes factors, and the fit cutoff frequency essentially matches with each other, confirming that the physical content of the model resides in the existence of a cutoff rather than its detailed functional form, and thereby validating the DEGWB as a robust model family.
Not every cutoff prescription is admissible, however. A hard cutoff is clearly incompatible with the PTA data and receives a very low Bayes factor (Table~\ref{tab:2}), and the Bayes factor for a DEGWB model without any spectral cutoff is suppressed by more than an order of magnitude, strongly disfavoring that scenario. The NG15 data thus simultaneously rule out unphysical cutoff prescriptions and establish the physical necessity of a cutoff itself in the nanohertz SGWB.
Figure~\ref{fig:spet} shows the fit of the DEGWB model to the NG15 free spectrum under Hellings--Downs correlations. The agreement is excellent across the full frequency range, providing robust evidence in favor of the equilibrium spectrum of Eq.~(\ref{yingbian}) as the dominant component of the nanohertz SGWB observed today.

The analysis of PTA data not only lends strong empirical support to our theoretical framework but also provides a precise empirical determination of the cutoff frequency $\mathcal{W}$. From Eq.~(\ref{dandao2}) in the Methods section, this frequency controls the boundary behavior of ballistic diffusion and, through the non-Markovian factor $b$, fixes the mass scale at which the scale-dependent screening encoded in $G_{\mathrm{eff}}(m)$ becomes observationally relevant. With $\mathcal{W}$ determined by PTA data alone, this mass scale becomes a prediction rather than a free parameter.

It should be emphasized that $m_{c}$ is not a sharp threshold. The quantity $c^{3}/(4G\mathcal{W})$ merely sets the reference mass at which $b = 1/2$, whereas the screening becomes physically visible only once $G_{\mathrm{eff}}$ departs from $G$ by a phenomenologically significant amount, say $b\sim 0.1$--$0.9$. Using the best-fit value $\log_{10}\mathcal{W}=-8.24^{+0.14}_{-0.35}$, this yields
\begin{equation}
m_{c}\sim 10^{12}\text{--}10^{14}\,M_{\odot},
\end{equation}
which overlaps with the characteristic mass enclosed within spheres of radius $\sim 8\,h^{-1}\,\mathrm{Mpc}$ at the cosmic mean matter density, $M_{8}\approx 1.88\times 10^{14}\,h^{-1}\,M_{\odot}$.
This is the scale at which the rms density contrast in the present-day Universe takes the value $\sigma_{8}\approx 0.81$, marking the boundary between the linear regime of cosmic structure growth on larger scales and the nonlinear, virialized regime on smaller scales~\cite{Planck2018,Tully2008,Scrimgeour2016}. 
The agreement of our PTA-calibrated critical mass with this empirically determined transition scale, with no free cosmological parameter and no tuning beyond the independently measured value of $G$, is unlikely to be accidental.

The physical mechanism behind this concordance is the scale-dependent screening encoded in $G_{\mathrm{eff}}(m)$. For structures with $m\ll m_{c}$, $G_{\mathrm{eff}}\approx G$ and standard gravitational clustering proceeds unimpeded, and density perturbations grow into virialized, nonlinear objects, such as galaxies, groups, and clusters. For structures with $m\gtrsim m_{c}$, the effective coupling is reduced, the external gravitational traction from surrounding matter is screened, and collapse is suppressed. 

In standard cosmology, the mass scale separating the linear and nonlinear regimes of structure growth is not a free input but is determined dynamically by the evolution of primordial density perturbations. Starting from the nearly scale-invariant scalar spectrum generated during inflation, $\mathcal{P}_{\mathcal{R}}(k) = A_{s}(k/k_{*})^{n_{s}-1}$, with amplitude $A_{s}\approx 2.1\times 10^{-9}$ and spectral index $n_{s}\approx 0.965$~\cite{Planck2018}, the late-time matter power spectrum follows by convolution with the transfer function $T(k)$ encoding radiation-era suppression of sub-horizon modes~\cite{BBKS1986,Baumann2022} and with the linear growth factor $D(t)$ controlled by the matter and dark-energy density parameters $\Omega_{m}$ and $\Omega_{\Lambda}$~\cite{Planck2018}. The present-day nonlinear scale $R_{\rm NL}(t_{0})$ is defined implicitly by $\sigma(R_{\rm NL},t_{0})=1$, where $\sigma(R,t)$ is the rms density contrast smoothed on comoving scale $R$. Combining this condition with the BBKS transfer function~\cite{BBKS1986,Baumann2022} yields the schematic closed form
\begin{equation}\label{R_NL}
R_{\rm NL}(t_{0}) \;\approx\; \left[\frac{2\pi^{2}\,\sigma^{2}}{A_{s}\,\Omega_{m}\,H_{0}^{2}\,\mathcal{I}(n_{s})}\right]^{1/(3-n_{s})},
\end{equation}
in natural units, where $\sigma \equiv \sigma(R_{\rm NL},t_{0})=1$ enforces the linear/nonlinear boundary condition and $\mathcal{I}(n_{s})$ is a dimensionless $\mathcal{O}(1)$ integral factor encoding the shape of the transfer function~\cite{BBKS1986,Baumann2022}. 
For the measured values this yields $R_{\rm NL}(t_{0})\sim (6$--$10)\,h^{-1}\,\mathrm{Mpc}$ and
\begin{equation}\label{M_NL}
M_{\rm NL}(t_{0}) \;=\; \frac{4\pi}{3}\,\bar\rho_{m}(t_{0})\,R_{\rm NL}^{3}(t_{0}) \;\sim\; (3\text{--}16) \times 10^{13}\,h^{-1}M_{\odot},
\end{equation}
where $\bar\rho_{m}(t_{0}) = 3\Omega_{m}H_{0}^{2}/(8\pi G)$ is the present-day mean matter density and $H_{0}$ is the Hubble constant~\cite{Planck2018}. 

The DEGWB framework provides an independent mass scale $m_{c} \sim c^{3}/(4G\mathcal{W})$, determined not by the growth of primordial perturbations but by the cutoff of the fluctuation-dissipation kernel. The two descriptions are complementary rather than mutually exclusive: they characterize the same physical boundary through distinct mechanisms, one kinematic, following the linear growth of primordial perturbations, and the other dissipative, following the fluctuation-dissipation closure of the SGWB-matter coupling. The empirical concordance $m_{c}\sim M_{\rm NL}$ admits a stronger reading, promoting it to a structural identification, which yields a direct relation between the DEGWB cutoff and the cosmological sector,
\begin{equation}\label{W_cosmo}
\mathcal{W} \;=\; \frac{c^{3}}{4G\,M_{\rm NL}}.
\end{equation}

Equation~(\ref{W_cosmo}) admits a transparent physical reading in terms of coherence interaction between the SGWB and the matter it couples to. A GW mode of frequency $\omega$ maintains coherence over a timescale $\sim 1/\omega$, and for such a mode to drive a coherent back-reaction in a bound mass $M$, the mass must be able to respond gravitationally within that window. The natural response time set by gravitational systems is their Schwarzschild light-crossing time,
\begin{equation}\label{tau_schw}
\tau_{\rm Schw}(M) \;\sim\; \frac{2GM}{c^{3}},
\end{equation}
which defines the characteristic interval for the internal gravitational state of the mass to register and re-radiate a perturbation. Coherent coupling therefore requires $\tau_{\rm Schw}(M)\lesssim 1/\omega$, or equivalently $\omega\lesssim c^{3}/(2GM)$. Applied to the critical structure scale $M_{\rm NL}$, this immediately yields a high-frequency coherence threshold coinciding with the DEGWB cutoff $\mathcal{W}=c^{3}/(4GM_{\rm NL})$. Modes above $\mathcal{W}$ oscillate faster than matter at the linear-to-nonlinear transition can respond, and their back-reaction is therefore suppressed; only components with $\omega\lesssim\mathcal{W}$ couple effectively to these structures over cosmological times. The cutoff of the equilibrium spectrum is in this sense not a regularization imposed by hand but a fundamental coherence threshold for SGWB-matter coupling, set by the Schwarzschild light-crossing time of the critical structure scale.

Through $M_{\rm NL}$, Eq.~(\ref{W_cosmo}) expresses $\mathcal{W}$ entirely in terms of late-time cosmological observables $(A_{s},n_{s},\Omega_{m},H_{0})$, which themselves carry the imprint of early-universe physics through the linear growth chain. $\mathcal{W}$ is therefore not a fundamental constant of gravity, nor a Planckian or trans-Planckian scale, but a derived quantity linking nanohertz stochastic gravitational waves to the primordial power spectrum, the matter content, and the expansion rate of the Universe. Eq.~(\ref{W_cosmo}) thus constitutes a cross-sector consistency test of the framework, that any independent future determination of $\mathcal{W}$, from LISA, Taiji, or next-generation pulsar-timing arrays, can be checked against the cosmological combination $c^{3}/(4GM_{\rm NL})$ computed from CMB and large-scale-structure data. A first-principles derivation, in which the equilibrium SGWB is dynamically sustained by the clustered matter that couples to it and $\mathcal{W}(t)$ co-evolves with $M_{\rm NL}(t)$ across cosmic history, lies beyond the scope of the present work and will be reported separately.

\section{Conclusion and Outlook}

In this work, we have proposed a new perspective on the relationship between the SGWB and cosmic matter evolution. By combining linearized general relativity with non-equilibrium statistical mechanics, we constructed a generalized Langevin framework in which the SGWB acts as a stochastic driver of celestial-system motion, while the gravitational radiation emitted by the resulting accelerated motion constitutes a back-reaction that returns energy to the background. 
The system of matter and SGWB therefore evolves toward a dynamical equilibrium characterized by a fluctuation-dissipation relation and a universal equilibrium strain spectrum, the DEGWB, with a high-frequency cutoff $\mathcal{W}$ that arises as a coherence threshold for SGWB-matter coupling.
The coupling to this background is strongly scale-dependent, that light, small systems remain essentially transparent to the SGWB, while structures whose mass approaches and exceeds the characteristic scale $m_{c}\sim c^3/(4G\mathcal{W})$ experience a dissipative screening that can be recast as a scale-dependent effective gravitational constant $G_{\mathrm{eff}}(m) = G/(1+4Gm\mathcal{W}/c^3)$.

Three findings support the framework. First, fitting the equilibrium spectrum to the NANOGrav 15-year dataset yields a Bayes factor of $48\pm 3.8$ relative to the SMBHB baseline, comparable to the leading SIGW scenario but achieved entirely within general relativity and the Standard Model, without invoking new physics. Second, the phenomenological mass range over which the screening becomes effective, $m_{c}\sim 10^{12}\text{--}10^{14}\,M_{\odot}$ derived from the PTA-calibrated cutoff frequency, overlaps with the empirically determined linear-to-nonlinear transition scale of cosmic structure at $\sim 8\,h^{-1}\,\mathrm{Mpc}$. Since $\mathcal{W}$ is fixed by PTA data alone, this second agreement is a genuine prediction rather than a fit. Third, promoting the concordance $m_{c}\sim M_{\rm NL}$ to a structural identification yields $\mathcal{W} = c^{3}/(4GM_{\rm NL})$, in which the DEGWB cutoff is expressed entirely in terms of the late-time nonlinear mass scale derived from $\Lambda$CDM linear growth. The $\Lambda$CDM and DEGWB descriptions of this boundary are complementary rather than mutually exclusive, characterizing the same mass scale through two distinct physical mechanisms: the kinematic growth of primordial perturbations in one case, and the dissipative closure of the SGWB-matter coupling in the other. The cutoff itself admits a transparent physical interpretation as a phase-coherence threshold: gravitational-wave modes above $\mathcal{W}$ oscillate faster than matter at the critical structure scale can respond on its Schwarzschild light-crossing time $\tau_{\rm Schw}(M_{\rm NL})=2GM_{\rm NL}/c^{3}$, and their back-reaction is thereby suppressed. Modes below $\mathcal{W}$ remain coherent over the response time and couple effectively over cosmic history. $\mathcal{W}$ is thereby not a fundamental or Planckian scale, nor a regularization imposed by hand, but a derived quantity expressing a fundamental coherence threshold for SGWB-matter coupling and linking nanohertz gravitational-wave observables to the primordial power spectrum, matter content, and expansion rate of the Universe.

\textit{Outlook.}---The scale-dependent screening mechanism predicts a distinctive phenomenological signature that can, in principle, be tested by ongoing and upcoming large-scale structure surveys. Because $G_{\mathrm{eff}}$ is reduced for aggregates whose mass approaches and exceeds $m_{c}$, the DEGWB model does not act as a uniform rescaling of the matter power spectrum; rather, it predicts a crossover, growth proceeds unimpeded on scales below $k_{*}\sim 0.1\text{--}0.3\,h/\mathrm{Mpc}$ corresponding to $m_{c}$, and is damped on larger scales.

One observational arena where this mass-localized screening could matter is the ongoing discussion of the $S_{8}$ tension between CMB and low-redshift probes of structure growth~\cite{Asgari2021,Wright2025,Abbott2026,Abdalla2022,Ghirardini2024,Bocquet2024}. The current picture is mixed across surveys, with recent analyses such as KiDS-Legacy~\cite{Wright2025} consistent with Planck and others maintaining modest residual tensions, and a definitive resolution will likely await independent high-precision measurements from Euclid~\cite{Euclid2025} and the Vera C.\ Rubin Observatory~\cite{LSST2019}. DEGWB should be regarded in this context not as a proposed resolution of the tension but as a diagnostic with a specific, PTA-anchored and mass-localized prediction, distinguishable in principle from uniform-rescaling modified-gravity alternatives. A quantitative computation of the DEGWB-modified growth equation and the corresponding matter power spectrum $P_{\mathrm{DEGWB}}(k,z)$, and its confrontation with current and forthcoming weak-lensing data, is deferred to a companion paper.

Beyond the $S_{8}$ question, 
the equilibrium picture developed here, in which the SGWB is not a passive fossil record but a dynamically coupled participant in cosmic matter evolution, invites a broader re-examination of the role of stochastic gravitational backgrounds in precision cosmology. In a sense, the homogeneity and isotropy of the Universe on the largest scales may be understood as a large-scale manifestation of the second law of thermodynamics, realized through the coupling between matter and the SGWB.

Confirmation or falsification of this picture will ultimately rest on the next generation of observations. Space-borne GW antennas such as LISA and Taiji will map the SGWB across frequencies inaccessible to PTAs, providing independent constraints on the shape of the equilibrium spectrum and its cutoff. Euclid and LSST will measure the growth of structure with percent-level precision over a wide range of scales, enabling direct tests of the scale-dependent prediction outlined above. Together, these experiments will determine whether the hidden equilibrium between stochastic gravitational waves and matter proposed here is a genuine feature of our Universe.

\begin{acknowledgments}
This work is supported by the National Key Research and Development Program of China (Grant No. 2025YFE0217300, Grant No. 2021YFC2201901), and the International Partnership Program of the Chinese Academy of Sciences (Grant No. 025GJHZ2023106GC).
\end{acknowledgments}

\appendix
\section{Methods}\label{sec:method}
\subsection*{\textbf{Coordinate systems.}}
In the rest frame of the SGWB, spacetime hypersurfaces allow unambiguous averaging of events over time and space.
This rest frame corresponds to a comoving coordinate system aligned with cosmic expansion, enabling SGWB properties, such as its temperature, to be treated as global, averaged quantities. During the late-time universe, where variations in the scale factor become negligible and the Hubble parameter approaches zero, the effects of cosmic expansion are minimal. Under these conditions, the SGWB rest frame can be approximated by Minkowski spacetime.
While our analysis is conducted within this fixed coordinate system, the equivalence principle ensures that physical conclusions remain coordinate-independent, provided that observable effects are ultimately expressed in a covariant form.
We adopt the Minkowski metric with the signature 
$(-1,1,1,1)$ and employ natural units where the speed of light $c$ is set to unity starting from this part.

\subsection*{Fluctuating force on test mass in SGWB}
\par
In our framework, both the fluctuating and dissipative forces fundamentally originate from gauge effects. For a localized gravitational wave passing through a specific spacetime point, it is indeed possible to choose an appropriate gauge in which these effects vanish. However, such a gauge choice is inherently local. That is, it is not possible to construct a global coordinate system in which the entire SGWB simultaneously exhibits no fluctuating or dissipative behavior.
Since stochastic dynamics and statistical analysis are intrinsically based on global statistical observables, the existence of fluctuation and dissipation, though gauge-dependent at the local level, acquires physical and statistical significance at the ensemble level. Thus, despite their gauge origin, these effects represent physically meaningful features of the SGWB when viewed through the perspective of statistical physics.

In the late-time Universe, we decompose the spacetime metric into the flat background ${\eta }_{\mu \nu }$ and the superposition of GW perturbations ${h}_{\mu \nu  }\left ( t,\mathbf{x}\right )$ and consider conformal expansion,
\begin{equation}
{g}_{\mu \nu }=a^2\left ({\eta }_{\mu \nu }+{h }_{\mu \nu }+\mathcal{O}\left ( {h}^{2}\right )\right ),
\end{equation}
where $a$ is the scale factor.
The motion of a test mass $m$ in a SGWB is described by the geodesic equation,
\begin{equation}
\frac{{d}^{2}{x}^{\mu }}{d{s }^{2}}+ {\varGamma }_{\rho \sigma  }^{\mu }\frac{d{x}^{\rho }}{{d s }}\frac{d{x}^{\sigma  } }{{d s }}=0,
\end{equation}
where ${x}^{\mu }=\{t,x^i\},\ (i=1,2,3)$ are the coordinates of the test mass in the cosmological rest frame of the SGWB, and $s$ is its proper time.

In slow motion limit, the 4-velocity of the test mass can be written as $u^{\mu}=\{1+\mathcal{O}({v}^{2}), v^i+\mathcal{O}({v}^{2})\}$, with $v^i=dx^i/dt$. 
Then the above geodesic equation along certain directions (take the $x^i$-direction as an example) can be written as
\begin{equation}
\label{cedi}
 \frac{{d}^{2}{x}^{i }}{d{t }^{2}}+ {\varGamma }_{\rho \sigma  }^{i}{v}^{\rho }{v}^{\sigma  }=0.
\end{equation}
We can further simplify this formula using such a relationship, $\mathcal{O}(h)\mathcal{O}(v) \ll \mathcal{O}(h)$.
Then, Eq.~ (\ref{cedi}) becomes
\begin{equation}
\label{lianluo}
\frac{d{v}^{i}}{dt}+2H{v}^{i}=-{\varGamma }_{00}^{i},
\end{equation}
where $H$ is the Hubble parameter.
Then, within the slow-motion limit, $m{\varGamma }_{00}^{i}$ will appear as the fluctuating force in the generalized Langevin equation along $x^i$-direction.
In the weak field approximation, we can write the connection component ${\varGamma }_{00}^{i}$ in Eq.~ (\ref{lianluo}) as
\begin{equation}
\label{gama}
{\varGamma }_{00}^{i}(t,\mathbf{x})=\frac{1}{2}({\eta }^{\lambda i}-{h}^{\lambda i} )(2{h}_{\lambda 0,0}-{h}_{00,\lambda })+H({\eta }^{\lambda i}-{h}^{\lambda i} ){h}_{\lambda 0}.
\end{equation}
When considering diffusion effects in the late stage of cosmic expansion, since the magnitude of the Hubble parameter $H$ is very small, terms involving $H$ can be omitted when writing down the stochastic partial differential equation.

For ${h}_{\mu \nu }\left ( t,\mathbf{x}\right )$, in the cosmological rest frame of the SGWB, we can decompose it as follows:~\cite {Allen1999, Thrane2013}
\begin{equation}
{h}_{\mu \nu }\left ( t,\mathbf{x}\right )=\frac{1}{ 2\pi }\int_{-\infty }^{\infty }d\omega \int_{-\infty }^{\infty }d{}^{3}\mathbf{k} {h}_{\mu \nu }(\omega ,\mathbf{k}){e}^{-i\omega t+i\mathbf{k}\cdot \mathbf{x}},
\end{equation}
where $\mathbf{k}$ is the three-dimensional wave vector and must satisfy the condition ${\omega }^{2}={k}^{2}$. 
For a stationary and uniform SGWB, we have,
\begin{equation}
\left \langle {h}_{\mu \nu }\left ( t,\mathbf{x}\right )\right \rangle{,}_{\lambda }=0
,\end{equation}
which leads to the conclusion that the means of the connection components should vanish, $\left \langle {\varGamma }_{\mu \nu }^{\rho }\left ( t\right )\right \rangle=0$. 
Here, $\left \langle …\right \rangle$ represents the result of statistical averaging.
From this, it follows that for a test mass moving at constant velocity, the stochastic influence of the SGWB, as described by the geodesic equation, statistically averages to zero. In contrast, the Langevin equation reveals that dissipative effects persist in the motion equation even after averaging the stochastic action. Thus, no dissipative effects can be derived directly from the geodesic equation.

The strain power spectrum density of ${h}_{\mu \nu }\left ( t,\mathbf{x}\right )$ is expressed as,
\begin{equation}
\left \langle {{h}^{\dagger}}_{\mu \nu }(\omega {}^{\prime},\mathbf{k}{}^{\prime}){h}_{\mu \nu }(\omega ,\mathbf{k})\right \rangle=\frac{1}{2}{\delta }^{3} ({\mathbf{k}}^{\prime}-\mathbf{k})\delta ({\omega }^{\prime}-\omega )S_{\mu \nu }(\omega ).
\end{equation}
At the same time, we apply the harmonic coordinate condition,
\begin{equation}
\overline{h}{}_{\mu \alpha }^{, \alpha }(t,\mathbf{x})=0,
\end{equation}
where,
\begin{equation}
\overline{h}{}_{\mu \nu }={h}_{\mu \nu }-\frac{1}{2}{\eta }_{\mu \nu }{h_{\lambda }}^{\lambda }.
\end{equation}
Using this condition, the degrees of freedom of the metric are reduced to six, corresponding to ${h}_{ij}$. By using the assumption of homogeneity of space, the six random variables ${h}_{ij}$ 
must satisfy identical statistical properties, meaning they share the same mean and higher-order moments. Consequently, their PSD function can be unified as
\begin{equation}
{S}_{ij}\left ( \omega \right )=S\left ( \omega \right ).
\end{equation}
Using Eq.~ (\ref{gama}), the Christoffel symbol for the $x^3$-direction is expressed as
\begin{equation}
\label{ga}
{\varGamma }_{00}^{3}\left ( \omega ,\mathbf{k}\right )=-i\left (\omega {h}_{30}+\frac{1}{2}{k}_{3}{h}_{00} \right ).
\end{equation}
Applying the third component of the harmonic coordinate condition,
\begin{equation}
\omega {\overline{h}}_{30}\left ( \omega ,\mathbf{k}\right )+{k}_{1} {\overline{h}}_{31}\left ( \omega ,\mathbf{k}\right )+{k}_{2} {\overline{h}}_{32}\left ( \omega ,\mathbf{k}\right )+{k}_{3} {\overline{h}}_{33}\left ( \omega ,\mathbf{k}\right )=0,
\end{equation}
we rewrite Eq.~ (\ref{ga}) as 
\begin{equation}
\label{sf}
{\varGamma }_{00}^{3}\left ( \omega ,\mathbf{k}\right )=i\left ( {k}_{1}{h}_{13}+{k}_{2}{h}_{23}\right )-\frac{i}{2}{k}_{3}\left ({h}_{11}+{h}_{22}-{h}_{33} \right ).
\end{equation}
The PSD function of the fluctuation force $F^3 $ along the $x^3$-direction, arising from the uniform and isotropic SGWB, is given by
\begin{equation}
{S}_{F^3}\left ( \omega \right )=\frac{11}{12}{m}^{2}S\left ( \omega \right ){\omega }^{2}.
\end{equation}
For the 3D case, the PSD function is $3{S}_{F^3}\left ( \omega \right )$, based on isotropy. By contracting ${h}_{\mu \nu }$ in the rest frame of the SGWB, we obtain a coordinate-independent strain power spectral ${S}_{h}\left ( \omega \right )$.

\subsection*{Gravitational dissipation}
In GR, gravitational radiation induces a back-reaction force on celestial bodies, altering their orbital dynamics. 
This is a type of motion of the test mass in the gravitational field other than geodesic motion, and the effects of this back-reaction force cannot be eliminated through statistical averaging.
To calculate the power of energy dissipation due to back-reaction force on the test mass in the SGWB, denoted as $\left \langle {P}_{\text{GD}}\right \rangle$, it is crucial to note its physical meaning. Specifically, $\left \langle {P}_{\text{GD}}\right \rangle$ represents the time-averaged energy transferred from the test mass to the surrounding background per unit time, where the "background" encompasses all cosmic constituents except for this test mass. The corresponding dissipation arises from gravitational radiation back-reaction, that a mass $m$ accelerating $\vec{a}$ through the SGWB emits metric perturbations described by the linearized Einstein field equations, carrying energy away at the rate
\begin{equation}\label{gonglv}
P_{\mathrm{\text{GD}}} = {4Gm^{2}|\vec{a}|^{2}}.
\end{equation}

To compute this back-reaction force, we adopt the Lorenz gauge and the weak-field approximation~\cite{Barack2019}.
Assuming that the test mass
$m$ is initially located near the origin, moving slowly along the $x^3$-axis with velocity $v$, the energy-momentum tensor for a point particle is given by:
\begin{equation}
{T }_{\mu \nu}\left ( \mathbf{x},t\right )=m{u}_{\mu }{u}_{\nu }\delta{}^{3} \left ( \mathbf{x}-\mathbf{z}(t)\right ),
\end{equation}
where $\mathbf{z}(t)$ is the trajectory of the test mass.
In this case, the perturbations ${h}_{\mu \upsilon }$ are described by the linearized Einstein field equations~\cite{Weinberg1972,Maggiore2007},
\begin{equation}
{\Box }^{2}{h}_{\mu \upsilon }=-16\pi G{\mathcal{S}}_{\mu \nu },
\end{equation}
where 
\begin{equation}
{\mathcal{S}}_{\mu \nu }={T}_{\mu \nu }-\frac{1}{2}{\eta }_{\mu \nu }T_{\lambda }\,^{\lambda }.
\end{equation}
One solution to this equation is the retarded solution,
\begin{equation}
{h}_{\mu \nu }(\mathbf{x},t)=4G\int {d}^{3}\mathbf{x}^{\prime}\frac{{\mathcal{S}}_{\mu \nu }(\mathbf{x}^{\prime},t-\left | \mathbf{x}-\mathbf{x}^{\prime}\right |)}{\left | \mathbf{x}-\mathbf{x}^{\prime}\right |},
\end{equation}
where $\mathbf{x}^{\prime}$ represents the position of the source and only takes values within the source distribution region.
At the location very far from the source, $\mathbf{x}\gg \mathbf{x}^{\prime}$, we replace $\left | \mathbf{x}-\mathbf{x}^{\prime}\right |$ with $r$, $(r=\left | \mathbf{x}\right |)$.

Then we work out the solutions of ${h}_{\mu \nu }$, and non-zero terms are ${h}_{00},{h}_{11},{h}_{22},{h}_{33}$ and ${h}_{03}$. 
${h}_{03}$ is written as
\begin{equation}
\begin{aligned}{h}_{03}(\mathbf{x},t)
 &=4G\int {d}^{3}{x}^{\prime}\frac{{T}_{03}({\mathbf{x}}^{\prime},t-r)}{r} \\ 
&=\frac{4G}{r}mv^*,
\end{aligned}
\end{equation}
where $v^*$ is the retarded velocity. If the test mass is isolated, it must satisfy ${{h}}_{03,0}(\mathbf{x},t)=0$. However, in our case, there will be an acceleration $a$ of the test mass with respect to the rest frame driven by the disturbances from the SGWB. 
As for other non-zero terms, their specific forms are written as:
\begin{equation}
\begin{aligned}{h}_{ 33}(\vec{x},t)
 & =4G\int {d}^{3}{x}^{\prime}\frac{{\mathcal{S}}_{33}({\vec{x}}^{\prime},t-\left | \vec{r}\right |)}{\left | \vec{r}\right |}\\ 
 & =\frac{2G}{r}m\left ( v{}^{*}\right ){}^{2}+\frac{2G}{r}m,
\end{aligned}
\end{equation}
\begin{equation}
{h}_{ 00}(\vec{x},t)=\frac{2G}{r}m\left ( v{}^{*}\right ){}^{2}+\frac{2G}{r}m,
\end{equation}
\begin{equation}
{h}_{ 11}(\vec{x},t)=\frac{2G}{r}m-\frac{2G}{r}m\left ( v{}^{*}\right ){}^{2},
\end{equation}
\begin{equation}
{h}_{ 22}(\vec{x},t)=\frac{2G}{r}m-\frac{2G}{r}m\left ( v{}^{*}\right ){}^{2}.
\end{equation}
Then, one has $h_{\mu\mu,0}\sim\left ( Gm v{}^{*}a{}^{*}/r\right ),\ (\mu=0,1,2,3)$. 
When the velocity $v$ is a small quantity, ${{h}}_{\mu\mu,0}\sim \mathcal{O}(v\,h_{03,0})$ is much smaller than ${{h}}_{03,0}$.

Once the specific form of  ${h}_{\mu \nu }$ is determined, the corresponding dissipated energy can be calculated using the average energy-momentum tensor of GWs. The energy dissipation rate per unit solid angle in a given direction is expressed as ~\cite{Weinberg1972}:
\begin{equation}
\label{dP}
\begin{aligned}\frac{dP}{d\Omega }(\mathbf{x },\omega)=
 &  \frac{{r}^{2}{\omega }^{2}}{32\pi G}[ {h}^{\mu \nu \dagger }(\mathbf{x},\omega){h}_{\mu \nu }(\mathbf{x },\omega)\\ 
 & \qquad \qquad \qquad \qquad -\frac{1}{2}\left | {h}_{\lambda }\,^{\lambda }\left ( \mathbf{x},\omega \right )\right |{}^{2}].
\end{aligned}
\end{equation}
Here, ${h}_{\mu \nu }(\mathbf{x},\omega)$ is the Fourier transform of ${h}_{\mu \nu }(\mathbf{x},t)$ at a fixed spatial position $\mathbf{x}$:
\begin{equation}
{h}_{\mu \nu }(\mathbf{x},t )=\int_{-\infty }^{\infty }{h}_{\mu \nu }(\mathbf{x},\omega ){e}^{i\omega t}d\omega.
\end{equation}
Eq.~ (\ref{dP}) can be rewritten as
\begin{equation}
\begin{aligned}\frac{dP}{d\Omega }(\mathbf{x },\omega)
 &=\frac{{r}^{2}}{32\pi G} [ {{h}_{,0}}^{\mu \nu \dagger }(\mathbf{x},\omega){{h}}_{\mu \nu ,0}(\mathbf{x },\omega) \\ 
 & \qquad \qquad \qquad \qquad  -\frac{1}{2}\left | {h}_{\lambda ,0}\,^{\lambda }\left ( \mathbf{x},\omega \right )\right |{}^{2} ].
\end{aligned}
\end{equation}
Based on the form of the average power, we calculate the total energy emitted per unit of solid angle emitted in direction $\mathbf{\hat{x}}$ (stands for the normal vector along $\mathbf{x}$) as follows:  
\begin{equation}
\frac{dE}{d\varOmega }(\mathbf{x})=2\pi  \int_{-\infty }^{\infty}\frac{dP}{d\Omega }(\mathbf{x },\omega)d\omega .
\end{equation}
Plug in the expression for $\frac{dP}{d\Omega }(\mathbf{x },\omega)$,
\begin{equation}
\begin{aligned}\frac{dE}{d\varOmega }(\mathbf{x})
 &= \frac{{r}^{2}}{16G}\int_{-\infty }^{\infty }d\omega [{{h}_{,0}}^{\mu \nu \dagger }(\mathbf{x},\omega){{h}}_{\mu \nu ,0}(\mathbf{x },\omega)\\ 
 & \qquad \qquad \qquad \qquad \qquad -\frac{1}{2}\left | {h}_{\lambda,0 }\,^{\lambda }\left ( \mathbf{x},\omega \right )\right |{}^{2}  ]\\ 
 &=\frac{{r}^{2}}{32\pi G }\int_{-\infty }^{\infty }dt [{{h}_{,0}}^{\mu \nu \dagger }(\mathbf{x},t){{h}}_{\mu \nu,0 }(\mathbf{x },t) \\ 
 & \qquad \qquad \qquad \qquad \qquad -\frac{1}{2}\left | {h}_{\lambda,0 }\,^{\lambda }\left ( \mathbf{x},t\right )\right |{}^{2}  ],
\end{aligned}
\end{equation}
where the second equal sign has used the Parseval's theorem:
\begin{equation}
\frac{1}{2\pi }\int_{-\infty }^{\infty}\left | {h}_{\mu \nu ,0}(t)\right |{}^{2}dt=\int_{-\infty }^{\infty}\left | {h}_{\mu \nu ,0}(\omega )\right |{}^{2}d\omega .
\end{equation}
Then we plug in the specific form of ${h}_{\mu \nu,0}$ for disturbed test masses. When evaluating the total energy emitted, we focus on the dominant contributions from the leading components, ${h}_{03,0}$ and ${h}_{30,0}$:
\begin{equation}
\frac{dE}{d\varOmega }(\mathbf{x})=\frac{{r}^{2}}{32\pi G }\int_{-\infty }^{\infty }\frac{32{G}^{2}}{{r}^{2}}{m}^{2}{a}^{*2}dt.
\end{equation}
The total power of energy dissipated by the test mass due to its back-reaction force is given by
\begin{equation}
\label{gonglv2}
{P}_{\text{GD}}=\frac{dE}{dt}=4 G{m}^{2}{a}^{2}.
\end{equation}
This result is analogous to the dissipation effect described by the Larmor formula for charged particles in a random electromagnetic field.

Dissipation effects also occur in the other two spatial directions. If the test mass undergoes accelerations along these axes, they contribute to the total energy dissipation. Consequently, the total power of energy dissipation is proportional to $\left | \vec{a}\right |{}^{2}$, as expressed in Eq.~(\ref{gonglv}).

Traditionally, the influence of gravitational waves on celestial systems has been described using the geodesic deviation equation. In this picture, as gravitational waves pass through spacetime, they induce only tidal deformations in celestial systems. The associated back-reaction—arising from the quadrupole moment variations induced by these deformations—is treated as a higher-order effect. Both the coupling between gravitational waves and celestial systems and the resulting back-reaction are thus regarded as perturbatively small. From this perspective, matter is effectively transparent to gravitational waves.

While the traditional framework is undoubtedly valid when describing local spacetime, it is not adequate for the physical picture and problems we aim to address. First, the standard geodesic deviation equation is intrinsically local and is not well suited to scenarios involving super-galactic scales or spacetimes with curvature gradients governed by stochastic processes. Second, the Langevin model we adopt is a phenomenological framework, constructed to describe the effective dynamics of a celestial system undergoing random motion. In the rest frame of the SGWB, the primary contribution to the celestial system’s motion comes from the geodesic equation defined in this coordinate system.
Due to the stochastic nature of the SGWB, the connection components ${\varGamma }_{00  }^{i}$ do not remain zero over time in any fixed global frame. Under the low-velocity approximation, $m{\varGamma }_{00  }^{i}$ thus plays the role of a fluctuating force, driving the random motion.

This perspective suggests that the impact of gravitational waves on celestial systems extends beyond tidal deformation. To achieve an equilibrium between the fluctuating force and the dissipative force (a requirement imposed by fluctuation-dissipation relations), a lower-order form of back-reaction must be included. This leads us to a picture where celestial systems and the SGWB form a dynamically coupled system, with mutual exchange of energy and information, rather than a nearly transparent influence from the gravitational field alone. This conceptual framework bears a strong analogy to the motion of a charged particle in a stochastic electromagnetic field, where dissipation arises from the radiation of electromagnetic energy. In our case, the back-reaction is likewise derived from the linearized Einstein field equations.

\subsection*{Generalized Langevin equation for test mass in SGWB}

We assume that test masses reach equilibrium with the SGWB after undergoing prolonged random motion. Then, we multiply both sides of the Eq.~ (\ref{gl}) by ${v}_{i}\left ( t\right )$,
\begin{equation}
\label{conserve}
\frac{m}{2}\frac{d{v}_{i}^{2}\left ( t\right )}{dt}=-m\int_{-\infty }^{\infty}\tilde{\gamma }\left ( t-{t}^{\prime}\right ){v}_{i}\left ( {t}^{\prime}\right ){v}_{i}\left ( t\right )d{t}^{\prime}+F_{i}\left ( t\right ){v}_{i}\left ( t\right ),
\end{equation}
where we have introduced the function $\tilde{\gamma }\left ( t\right )=\Theta \left ( t\right )\gamma \left ( t\right )$, and $\Theta \left ( t\right )$ is the Heaviside function which vanish for $t< 0$. 
We then average the resulting expression over time. $\left \langle F_i\left ( t\right )v_i\left ( t\right )\right \rangle$  represents the average work done on the Brownian test mass in a specific direction.
Given $t\gg 0$, one has $m\left \langle v{}^{2}\right \rangle/2={E}_{k}=constant$, so the left side of Eq.~ (\ref{conserve}) equals 0. The right side represents that the average dissipation power equals the average power of work done by the background.
Then, we obtain the dissipated power, $\left \langle {P}_{\text{GD}}\right \rangle$,
 \begin{equation}
\begin{aligned}\left \langle {P}_{\text{GD}}\right \rangle
 &=m\int_{-\infty }^{\infty}\tilde{\gamma }\left ( t-{t}^{\prime}\right ) \left \langle \vec{v}\left ( {t}^{\prime}\right )\cdot\vec{v}\left ( t\right )\right \rangle d{t}^{\prime} \\ 
 & =m\int_{-\infty }^{\infty}\tilde{\gamma }\left ( \tau \right ){\kappa }_{\left | \vec{v}\right |} \left ( \tau \right )d\tau ,
\end{aligned}
\end{equation}
where ${\kappa }_{\left | \vec{v}\right |} \left ( \tau \right )$  is the autocorrelation function of velocity. ${\kappa }_{\left | \vec{v}\right |}  \left ( \tau \right )$ is an even function, so we make an even extension of $\tilde{\gamma }\left ( \tau \right )$. $\left \langle {P}_{\text{GD}}\right \rangle$ becomes
\begin{equation}
\left \langle {P}_{\text{GD}}\right \rangle=\frac{m}{2}\int_{-\infty }^{\infty}{\gamma }\left ( \tau \right ){\kappa }_{\left | \vec{v}\right |}  \left ( \tau \right )d\tau .
\end{equation}
Next, we take the Fourier transform of ${\gamma }\left ( \tau \right )$ and ${\kappa }_{\left | \vec{v}\right |}  \left ( \tau \right )$ simultaneously, expressing the energy dissipation in the frequency domain,
\begin{equation}
\begin{aligned}\left \langle {P}_{\text{GD}}\right \rangle
 & =\frac{m}{2}\int_{-\infty }^{\infty}d\tau d{\omega } {d\omega' }\frac{{\gamma }_{f}\left ( {\omega }\right ){S}_{\left | \vec{v}\right |} \left ( {\omega' }\right )}{(2\pi)^{2}}{e}^{i\left ( {\omega }+{\omega' }\right )\tau }\\ 
 & =\frac{m}{2}\left ( \frac{1}{2\pi }\right )\int_{-\infty }^{\infty}d\omega {\gamma }_{f}\left ( \omega \right ){S}_{\left | \vec{v}\right |} \left ( \omega \right ),
\end{aligned}
\end{equation}

Based on the generalized Langevin model, the relationship between the PSD function of velocity and the PSD function of the fluctuating force is written as follows ~\cite{Pottier2010} :
\begin{equation}
\label{Sv}
{S}_{\left | \vec{v}\right |}\left ( \omega \right )=\frac{1}{{m}^{2}}\frac{1}{\left | \gamma \left ( \omega \right )-i\omega \right |{}^{2}}{S}_{\left | \vec{F}\right |}\left ( \omega \right ),
\end{equation}
where $ \gamma \left ( \omega \right )$ is the Fourier–Laplace transform of $ \gamma \left ( t\right )$:
\begin{equation}
\gamma \left ( \omega \right )=\int_{0}^{\infty }\gamma \left ( t\right ){e}^{i\omega t}dt.
\end{equation}
The relation between ${\gamma }_{f}\left ( \omega \right )$ and $ \gamma \left ( \omega\right )$ reads
\begin{equation}
{\gamma }_{f}\left ( \omega \right )=2\mathfrak{R}e \gamma \left ( \omega \right ).
\end{equation}
We notice that $\left \langle {P}_{\text{GD}}\right \rangle$ can also be derived directly through GW radiation back reactions in the previous subsection and  be written as the function of $\left \langle {\vec{a}}^{2}\right \rangle$ as in Eq.~ (\ref{gonglv}),  and $a$, being the mean square derivative of $v$, has the PSD function according to Eq.~ (\ref{Sv}) 
\begin{equation}
{S}_{\left | \vec{a}\right |}\left ( \omega \right )=\frac{1}{{m}^{2}}\frac{{\omega }^{2}}{\left | \gamma \left ( \omega \right )-i\omega \right |{}^{2}}{S}_{\left | \vec{F}\right |}\left ( \omega \right ).
\end{equation}
Therefore, one has
\begin{equation}
\label{PGD}
\begin{aligned}\left \langle {P}_{\text{GD}}\right \rangle
 &=4 G{m}^{2}\left \langle {\vec{a}}^{2}\right \rangle \\ 
 & =4 G{m}^{2}\frac{1}{2\pi }\int_{-\infty }^{\infty }\frac{1}{{m}^{2}}\frac{{\omega }^{2}}{\left | \gamma \left ( \omega \right )-i\omega \right |{}^{2}}{S}_{\left | \vec{F}\right |}\left ( \omega \right )d\omega .
\end{aligned}
\end{equation}
Comparing the two expressions of the energy distribution of $⟨P_{\text{GD}}⟩$ in the frequency domain results in Eq.~(\ref{haosan}).
Using the fluctuation-dissipation relation,
$\mathfrak{R}e \gamma\left ( \omega  \right )={S}_{\left | \vec{F}\right |}\left ( \omega  \right )/{6mkT},$
this constraint translates directly into restrictions on the strain power spectrum of the SGWB.
The temperature $T$ here is defined by the final average kinetic energy of the celestial system undergoing random motion under the significant influence of the SGWB, $\left | E\right |=3kT/2$. From Eq.~(\ref{haosan}) and fluctuation-dissipation relation, we have:
\begin{equation}
\label{temp}
T=\frac{1}{G{k}_{B}}\frac{11A^2}{480}.
\end{equation}
This temperature is independent of the mass of the celestial systems and is instead an intrinsic property of the SGWB itself.
From Eq.~(\ref{yingbian}), we can deduce the corresponding energy density distribution of the SGWB in the frequency domain:
\begin{gather}
\frac{d{\rho }_{GW}}{d\omega }=\frac{{\omega }^{2} }{32{\pi }^{2}G}{S}_{h}\left ( \omega \right ),\\
{\Omega }_{GW}\left ( f\right )=\frac{1}{12\pi {H}_{0}^{2}}{\omega }^{3}{S}_{h}\left ( \omega \right ).
\end{gather}
Interestingly, this distribution takes the form of the well-known Rayleigh–Jeans law.
The derivation of the Rayleigh–Jeans law is grounded in the classical equipartition theorem and provides an excellent approximation to the Planck blackbody spectrum at low frequencies.

From the Eq.~(\ref{haosan}), $\gamma(t)$ can be recovered:
\begin{equation}
\label{haosan2}
\gamma \left ( t\right )=4Gm{\mathcal{W} }^{2}\left [ \delta \left ( t\right )-\mathcal{W} e^{-\mathcal{W} \left | t\right | }\right ].
\end{equation}
Then, the mean and the second moment of the velocity of the celestial system can be expressed as follows:
\begin{gather}
\left \langle \vec{v}\left ( t\gg \tau  \right )\right \rangle=b\vec{v}\left ( 0\right ),
\\
\label{dandao2}
\left \langle \left | \vec{v}\right |{}^{2}\left ( t\gg \tau  \right )\right \rangle=\frac{3{k}_{B}T}{m}+{b}^{2}\left [ {\left | \vec{v}\left ( 0\right )\right |}^{2}-\frac{3{k}_{B}T}{m}\right ].
\end{gather}

\subsection*{Cut-off Factor}

We introduce a cutoff factor, ${f}_c$, which is defined such that it equals 1 for frequencies up to a large cutoff value $\mathcal{W}$ and transitions to zero beyond this point. The specific form of the cutoff factor does not influence the results of long-term evolution; the only relevant parameter affecting the physical outcomes is the cutoff frequency, ${\mathcal{W}}$. A convenient form that satisfies this condition is ~\cite{Ford1988}
\begin{equation}
{f}_c=\frac{{\mathcal{W} }^{2}}{\omega {}^{2}+{\mathcal{W} }^{2}}.
\end{equation}
Then, we can multiply $\mathfrak{R}e \gamma \left ( \omega \right )$ by ${f}_c$
\begin{equation}
\label{wentai}
\mathfrak{R}e \gamma \left ( \omega \right )=4 Gm{\omega }^{2}{f}_{c}.
\end{equation}
If Eq.~ (\ref{haosan}) is assumed to hold across all frequencies, this corresponds to the limit $\mathcal{W} \to \infty $, where ${f}_{c}=1$ for all frequencies.
In this case, we have:
\begin{equation}
\gamma \left ( t\right )=-4Gm\ddot{\delta }\left ( t\right ).
\end{equation}
Then, the Fourier–Laplace transform of $ \gamma \left ( t\right )$ is
\begin{equation}
\begin{aligned}\gamma \left ( \omega \right )
 & =\int_{0}^{\infty }-4Gm\ddot{\delta }\left ( t\right ){e}^{i\omega t}dt\\ 
 & =4Gm{\omega }^{2}
\end{aligned}
\end{equation}
According to the fluctuation-dissipation theorem, $\mathfrak{R}e \gamma\left ( \omega \right )$ and ${S}_{F}\left ( \omega  \right )$ differ only by a constant factor $6mkT$. Substitute ${S}_{F}\left ( \omega  \right )$ and $\gamma \left ( \omega \right )$ into the Eq.~ (\ref{PGD}). One will find that $\left \langle {P}_{\text{GD}}\right \rangle$ becomes infinity in this case. This is not by physical reality. So, $\mathcal{W} \to \infty $ represents a condition that does not conform to physical facts, and the introduction of a cut-off factor ${f}_{c}$ is necessary. 
Furthermore, the presence of a cutoff factor is also necessary to ensure that the power of the background energy density does not diverge.

\begin{table*}
\caption{Prior distributions for each model in Bayesian analysis}
\label{tab:3}
\centering
    \begin{tabular}{p{2cm}<{\centering}p{8cm}<{\centering}p{4cm}<{\centering}}
        \toprule
        \toprule
        Parameter   &  Description & Prior distribution \\
        \midrule
        \multicolumn{3}{c}{\textbf{DEGWB}}\\
        $S$            & The DEGWB amplitude   & log-Uniform $[-12,-8]$ \\
        $\mathcal{W}$  & Cut-off frequency         & log-Uniform $[-10,-5]$ \\
        $^*n$          & Power index (for smooth cut-off)               & Uniform $[0,5]$\\
        \midrule
        \multicolumn{3}{c}{\textbf{SMBHB}}\\
        $A$       & The SMBHB signal amplitude        & log-Uniform $[-18,-12]$ \\
        \midrule
        \multicolumn{3}{c}{\textbf{Scalar-induced Gravitational Waves(SIGW)~\cite{Afzal2023}}}\\
        $A$            & Scalar amplitude     & log-Uniform $[-3,1]$ \\
        $f_{min}$[Hz]  & Lower frequency      & log-Uniform $[-11,-5]$ \\
        $f_{max}$[Hz]  & Upper frequency      & log-Uniform $[-11,-5]$ \\
        \midrule
        \multicolumn{3}{c}{\textbf{Inflationary Gravitational Waves (IGW)~\cite{Afzal2023}}}\\
        $T_{rh}/$[GeV] & Reheating temperature  & log-Uniform $[-3,-3]$ \\
        $r$      & Tensor-to-scalar ratio      & log-Uniform $[-40,0]$ \\
        $n_t$    & Tensor spectral index       & Uniform $[0,6]$ \\
        \midrule
        \multicolumn{3}{c}{ \textbf{First-order phase transitions (FOPT)~\cite{Xue2021}}} \\
        $\alpha$ & Phase transition strength     & log-Uniform $[-5,-2]$ \\
        $T_*/$[GeV]    &  The phase transition temperature        & Uniform $[-6,2]$ \\
        $\beta/H_*$    & $H_*$ is the Hubble parameter at $T_*$ and  $\beta$ is the inverse duration of the phase transition   & log-Uniform $[-6,0]$ \\
        \midrule
        \multicolumn{3}{c}{\textbf{Domain Walls (DW)~\cite{Afzal2023}}} \\
        $a_{*}$        & Energy fraction in domain walls     & Uniform $[-3,0]$ \\
        $T_{*}/$[GeV]  & Transition temperature              & log-Uniform $[-4,-4]$ \\
        $b$            & High-frequency slope                & Uniform $[0.5,1]$ \\
        $c$            & Spectral shape width                & Uniform $[0.3,3]$ \\
        \bottomrule
        \bottomrule
    \end{tabular}
\end{table*}

\subsection*{The analysis of NG15 dataset}
In this section, we briefly introduce the SGWB models used for comparative analysis earlier. These theories have long been mainstream in the field and have received widespread attention~\cite{Afzal2023, Xue2021,Ellis2024source,Cai2023limits,2023Searching}. In the PTA band, the energy density spectrum of SIGWs can be expressed as:
\begin{equation}
    \Omega_{\mathrm{GW}}^{\mathrm{ind}}(f) = \Omega_{r}\left(\frac{g_{*}(f)}{g_{*}^{0}}\right)\left(\frac{g_{*, s}^{0}}{g_{*, s}(f)}\right)^{4 / 3} \bar{\Omega}_{\mathrm{GW}}^{\mathrm{ind}}(f).
\end{equation}
Here, $\Omega_r/g_*^0 \simeq 2.72 \times 10^{-5}$ is the current radiation energy density per unit critical density. $g_{*,s}^0 \simeq 3.93$ is the effective number of relativistic degrees of freedom contributing to the radiation entropy today. $g_{*}(f)$ and $g_{*, s}(f)$ describe the degrees of freedom when the gravitational waves re-enter the Hubble horizon. In the radiation-dominated era, the energy density spectrum of SIGWs can be further expressed as:
\begin{equation}
    \bar{\Omega}_{\mathrm{GW}}^{\mathrm{ind}}(f) = \int_{0}^{\infty}d\nu\int_{|1-\nu|}^{1+\nu}du\ \mathcal{K}(u,\nu)\mathcal{P}_{R}(uk)\mathcal{P}_{R}(\nu k),
\end{equation}
\begin{equation*}
    \begin{aligned}
\mathcal{K}(u,\nu) & =\frac{3(4\nu^2-(1+\nu^2-u^2)^2)^2(u^2+\nu^2-3)^4}{1024u^8\nu^8} \\
 & \times[\left(\ln\left|\frac{3-(u+\nu)^2}{3-(u-\nu)^2}\right|-\frac{4u\nu}{u^2+\nu^2-3}\right)^2 \\
 &
 +\pi^2\Theta(u+\nu-\sqrt{3})],
\end{aligned}
\end{equation*}
where $k=2\pi a_0 f$, and $a_0$ denotes the present value of the cosmological scale factor.
The kernel function $\mathcal{K}(u,\nu)$ reflects the second-order nonlinear coupling effect. For Gaussian-shaped SIGWs, we have:
\begin{equation}
\mathcal{P}_{R}(k)=\frac{A}{\sqrt{2 \pi} \Delta} \exp \left[-\frac{1}{2}\left(\frac{\ln k-\ln k_{*}}{\Delta}\right)^{2}\right] .
\end{equation}
Therefore, the SIGW model has three relevant parameters: the characteristic scale $k_{*}$ (in units of $\text{Mpc}^{-1}$), the peak amplitude $A$, and the width $\Delta$.

The energy density spectrum of IGWs can be expressed as:
\begin{equation}
    \Omega_{\mathrm{GW}}^{\inf}(f) = \frac{\Omega_{\mathrm{r}}}{24}\left(\frac{g_{*}(f)}{g_{*}^{0}}\right)\left(\frac{g_{*, s}^{0}}{g_{*, s}(f)}\right)^{4 / 3} \mathcal{P}_{\mathrm{t}}(f) T(f),
\end{equation}
where the inflationary dynamics generate the primordial tensor power spectrum:
\begin{equation}
    \mathcal{P}_{\mathrm{t}}(f) = r A_s \left( \frac{f}{f_{\mathrm{CMB}}} \right)^{n_t}.
\end{equation}
Here, \(r\) and \(A_s = 2.10 \times 10^{-9}\) represent the tensor-to-scalar ratio and the amplitude of the primordial scalar power spectrum, respectively. The main parameters of the IGW model include the reheating temperature \(T_{\mathrm{rh}}\), the tensor-to-scalar ratio \(r\), and the tensor spectral index \(n_t\).

In this paper, we only consider the contribution of sound waves during the First-Order Phase Transition (FOPT) to the energy density of the Stochastic Gravitational Wave Background (SGWB):
\begin{equation}
\begin{aligned}
h^{2}\Omega_{\mathrm{sw}}(f) & = 2.65 \times 10^{-6} v_{w} \left( \frac{H_{*}}{\beta} \right) \left( \frac{\kappa \alpha}{1 + \alpha} \right)^{2} \left( \frac{100}{g_{*}} \right)^{1/3} \\
& \times \left[ \left( \frac{f}{f_{\mathrm{sw}}} \right)^3 \left( \frac{7}{4 + 3 \left( \frac{f}{f_{\mathrm{sw}}} \right)^2} \right)^{7/2} \right] \\
& \times \min \left[ 1, (8\pi)^{1/3} v_{w} \left( \frac{H_{*}}{\beta} \right) \left( \frac{3}{4} \frac{\kappa \alpha}{1 + \alpha} \right)^{-1/2} \right],
\end{aligned}
\end{equation}
where \(g_{*}\) is the effective number of relativistic degrees of freedom; \(\kappa\) is the fraction of latent heat converted into plasma kinetic energy; \(v_{w}\) is the bubble wall velocity (taken as 1 in this paper); \(\alpha\) is the phase transition strength, which mainly determines the amplitude of the gravitational wave spectrum; \(\beta\) is the inverse of the phase transition duration (or the phase transition rate parameter); \(H_*\) is the Hubble parameter at the phase transition temperature \(T_*\). In the limit of $v_w \to 1$, $\kappa$ is uniquely determined by the phase transition strength $\alpha$. Thus, the FOPT model has three relevant parameters: phase transition strength \(\alpha\), phase transition temperature \(T_*\), and \(\beta/H_*\).

In the PTA band, the energy density spectrum of DWs can be expressed as:
\begin{equation}
h^2 \Omega_{\mathrm{GW}}(f) = \frac{3}{32\pi} \mathcal{D} \tilde{\epsilon} \alpha_{*}^{2} \mathcal{S}(f/f_{p}),
\end{equation}
where \(\mathcal{D} \approx 1.67 \times 10^{-5}\) is the dilution factor, \(\tilde{\epsilon} = 0.7\) is the efficiency coefficient obtained from numerical simulations, and \(\alpha_*\) is the fraction of the total energy density stored in domain walls at temperature \(T_*\). The gravitational wave background is primarily dominated by the emission before the decay of the domain wall network, so the typical frequency of the emitted gravitational waves is determined by the Hubble radius at the time of emission. After redshifting, the peak frequency of the GW background generated by DWs is expected to be:
\begin{equation}
f_p = 1.14 \, \mathrm{nHz} \left( \frac{10.75}{g_{*,s}} \right)^{1/3} \left( \frac{g_*}{10.75} \right)^{1/2} \left( \frac{T_*}{10 \, \mathrm{MeV}} \right).
\end{equation}
For the spectral shape \(\mathcal{S}(x)\), we adopt the following construction:
\begin{equation}
\mathcal{S}(x) = \frac{(a + b)^c}{(b x^{-a/c} + a x^{b/c})^c},
\end{equation}
where \(a = 3\).

Thus, the DW model has four relevant parameters: energy fraction in domain walls \(\alpha_*\), eransition temperature \(T_*\), high-frequency slope \(b\), and spectral shape width \(c\).

The prior ranges for the calculation of the BF of the parameters in the aforementioned models are presented in Table \ref{tab:3}.

\bibliographystyle{apsrev4-2}
\bibliography{Refarxiv}

\end{document}